\documentclass[aps,floats,twocolumn,preprintnumbers]{revtex4}
\usepackage{epsfig}
\usepackage{amsmath}
\usepackage{amssymb}
\usepackage{amsthm}
\usepackage{graphics}
\usepackage{float}
\usepackage{graphicx} 

\usepackage{pstricks}
\usepackage{pst-plot}
\usepackage{pst-grad}
\usepackage{pst-eps}

\newcommand{\bea}{\begin{eqnarray}}
\newcommand{\eea}{\end{eqnarray}}
\newcommand{\beq}{\begin{equation}}
\newcommand{\eeq}{\end{equation}}

\DeclareMathAlphabet{\mathpzc}{OT1}{pzc}{m}{it}

\begin{document}

\title{Casimir Effect on the brane}
\author{Antonino Flachi}
\email{flachi@yukawa.kyoto-u.ac.jp}
\affiliation{Yukawa Institute for Theoretical Physics, Kyoto University, Kyoto 606-8502, Japan}
\author{Takahiro Tanaka}
\email{tanaka@yukawa.kyoto-u.ac.jp}
\affiliation{Yukawa Institute for Theoretical Physics, Kyoto University, Kyoto 606-8502, Japan}
\preprint{YITP-09-28}

\begin{abstract}
We consider the Casimir effect between two parallel plates localized on a brane. We argue that in order to properly compute the contribution to the Casimir energy due to any higher dimensional field, it is necessary to take into account the localization properties of the Kaluza-Klein modes. When the bulk field configuration is such that no massless mode appears in the spectrum, as, for instance, when the higher dimensional field obeys twisted boundary conditions across the branes, the correction to the Casimir energy is exponentially suppressed. When a massless mode is present in the spectrum, the correction to the Casimir energy can be, in principle, sizeable. However, when the bulk field is massless and strongly coupled to brane matter, the model is already excluded without resorting to any Casimir force experiment. The case which is in principle interesting is when the massless mode is not localized on the visible brane. We illustrate a method to compute the Casimir energy between two parallel plates, localized on the visible brane, approximating the Kaluza-Klein spectrum by truncation at the first excited mode.  
We treat this case by considering a piston-like configuration and introduce a small parameter, $\varepsilon$, that takes into account the relative amplitude of the zero mode wave function on the visible brane with respect to the massive excitation. 
We find that the Casimir energy is suppressed by two factors: at lowest order in $\varepsilon$, the correction to the Casimir energy comes entirely from the massive mode and turns out to be exponentially suppressed; the next-to-leading order correction in $\varepsilon$ follows, instead, a power-law suppression due to the small wave function overlap of the zero-mode with matter confined on the visible brane. Generic comments on the constraints on new physics that may arise from Casimir force experiments are also made.
\end{abstract}
\maketitle
\vspace{2mm}

\section{Introduction}
In 1948 Casimir predicted the presence of an attractive force, of
nongravitational origin, between two parallel, neutral, perfectly
conducting plates \cite{casimir}. This phenomenon is understood as a
consequence of the non-trivial dependence of the energy of 
vacuum fluctuation structure on the macroscopic boundary conditions. 
A first principle calculation of the quantum vacuum pressure, $P_C$,
obtained by summing all the zero-point electromagnetic fluctuations,
predicts
\bea
P_C = {\hbar c \pi^2 \over 240 \ell^4} 
\approx 
1.3 \times 10^{-27} \left({\ell\over 1\mbox{m}}\right)^{-4}
\mbox{N m$^2$}~,
\eea
with $\ell$ being the distance between the plates. The experimental
verification of the Casimir effect has been pursued for various decades
\cite{report1,report2}. For the specific geometry of a sphere close to a
plane wall, the force has been measured unambiguously, up to now, with
an accuracy of $1\%$ \cite{lam,moh,Decca1}. The case of two parallel
plates, originally studied by Casimir,  was initially examined by
Sparnay, whose experimental results were affected by a large systematic
error and thus not conclusive \cite{sparnay}. Only comparatively
recently, the Casimir force between two parallel plates has been
successfully measured for a plate separation of $0.5-3.0~\mu$m
\cite{Bressi}. 

The chance to investigate experimentally the deformations of the vacuum
opened up the possibility to explore new physical phenomena at
microscopic scales \cite{Decca2}. This triggered many people to study
how the Casimir force changes when additional degrees of freedom, other
than the electromagnetic one, are present. In particular, during the
past couple of years considerable activity has concentrated in working
out modifications to the Casimir force in the presence of extra
dimensions (in some cases in conjuction with recent work on Casimir
pistons \cite{piston}) \cite{ed1,ed2,ed22,ed3,ed4,ed5,ed6,ed7,ed8}. 

In the braneworld scenario, gravity propagates throughout the bulk
space. On macroscopic length scales ($\agt 0.1 mm$), 
a massless graviton zero mode reproduces ordinary four dimensional general
relativity with the aid of an appropriate stabilization
mechanism~\cite{gt}. 
The Standard Model (SM) fields are, instead, localized on the
visible brane. These features are necessary to make the model compatible with
observations. Higher dimensional bulk fields are also expected to
exist from the view point of realistic model construction of brane
worlds. As in the 
standard Kaluza-Klein models with homogeneous and compact extra
dimensions, a bulk field is perceived on the visible brane as a tower of four
dimensional fields. 
The presence of bulk fields is also a
necessary ingredient to stabilize the moduli (distance between the
branes, for some examples see \cite{garr,ft,ftm}).

Here, we wish to examine the corrections to the Casimir force arising
from the presence of bulk fields. The Casimir effect is an 
{\it experimentally well tested physical phenomenon}, which can be
understood in terms of the vacuum fluctuations of the electro-magnetic
field. 
If there are large extra dimensions, their effect must be small so that
the model is compatible with current experiments. 
In scenarios with extra dimensions, there is a degree of model
dependence, which is accompanied by a non-trivial dependence of the
Casimir force on the geometry of the experimental apparatus. However, it
is easy to pin down what are the main features of higher dimensional
models that may in principle produce non-trivial modifications to the
Casimir force. 

In order to have sizeable low energy
effects, extra dimensions should be large enough. In this case, for phenomenological consistency (without resorting to Casimir force experiments), bulk fields
should be either weakly coupled to the SM or massive enough
in the four dimensional sense ($\agt$ TeV), to avoid
deviations from the SM predictions. 
Intuitively, therefore, one would expect the corrections to the Casimir force due
to higher dimensional fields also to be small. 
However, the Casimir force arises as a non-trivial effect of the vacuum
fluctuation, and hence may delicately depend on the renormalization. This
makes the issue of analyzing the corrections interestingly
subtle. 

It is the aim of this paper to reconsider the role of the corrections to
the Casimir force between the ideal system of two parallel plates in 
the context of viable braneworld models.

\section{Boundary conditions for Casimir force experiments}

\begin{figure}[t]
\includegraphics[height=5.25cm]{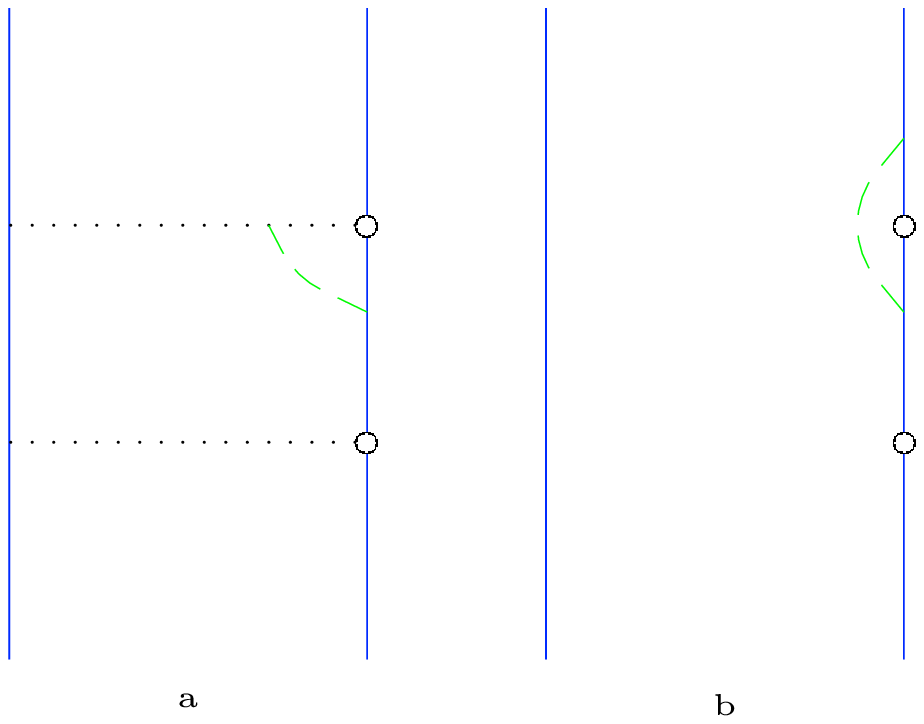}
   \caption{The figure schematically shows the two boundary conditions discussed in the paper. The continuous (blue) lines represent the branes, the circles represent the physical plates on the brane, and the dotted (black) lines signify that these plates are extended into the extra dimensions. The dashed (green) lines illustrates the propagation of a higher dimensional field. In the left hand panel (a), higher dimensional fields are confined in a `higher dimensional box' and boundary conditions are imposed on each mode. In the right hand pannel (b), the plates are physical (localized on the brane) and higher dimensional field can `by-pass' them, propagating through the bulk, even in the perfectly reflecting case.}
\label{figura}
\end{figure}

First, we wish to discuss the boundary conditions at the
plates in the braneworld setup. 
When there are no branes and the fields propagate in higher
dimensions, various examples have been examined in
Refs.~\cite{ed4,ed6}. However, the situation changes in the presence of
branes. In general, when the spacetime is a product space, the wave
function of a bulk field can be decomposed as
$$
\Phi= \sum_n \varphi_n(x_\mu) f_n(y)~,
$$
where $y$ represents coordinate(s) of the extra dimension(s). 
The four dimensional
effective action, assuming four dimensional Minkowski background spacetime, 
will take the standard tower-like form: 
\bea
S=-{1\over 2}\sum_n \int d^4x\, \varphi_n\left(\Box +m_n^2\right)\varphi_n~.
\label{efrsi}
\eea
The wave function $f_n(y)$ may or may not be peaked around the visible 
brane on which the SM fields are localized. 
The plates for which we measure the Casimir force
are composed of the SM fields, and hence they 
are localized on the brane. 
Technically it is difficult to impose such boundary conditions exactly. 
One possibile approximation is to
impose the boundary conditions on the respective excitation modes
independently. This corresponds to imposing the boundary conditions as 
if the plates were extended uniformly into the directions of
extra-dimensions. In other words, the scalar wave function is 
confined inside a ``higher dimensional box'' (see Fig.~\ref{figura}-a). 
This approximation will definitely over-estimate the Casimir force. 
We will discuss this case in the next section to show that 
the Casimir force is simply given by the sum of 
contributions from respective modes and 
is exponentially suppressed when all the modes are massive. 

\begin{figure}
\includegraphics[height=5cm]{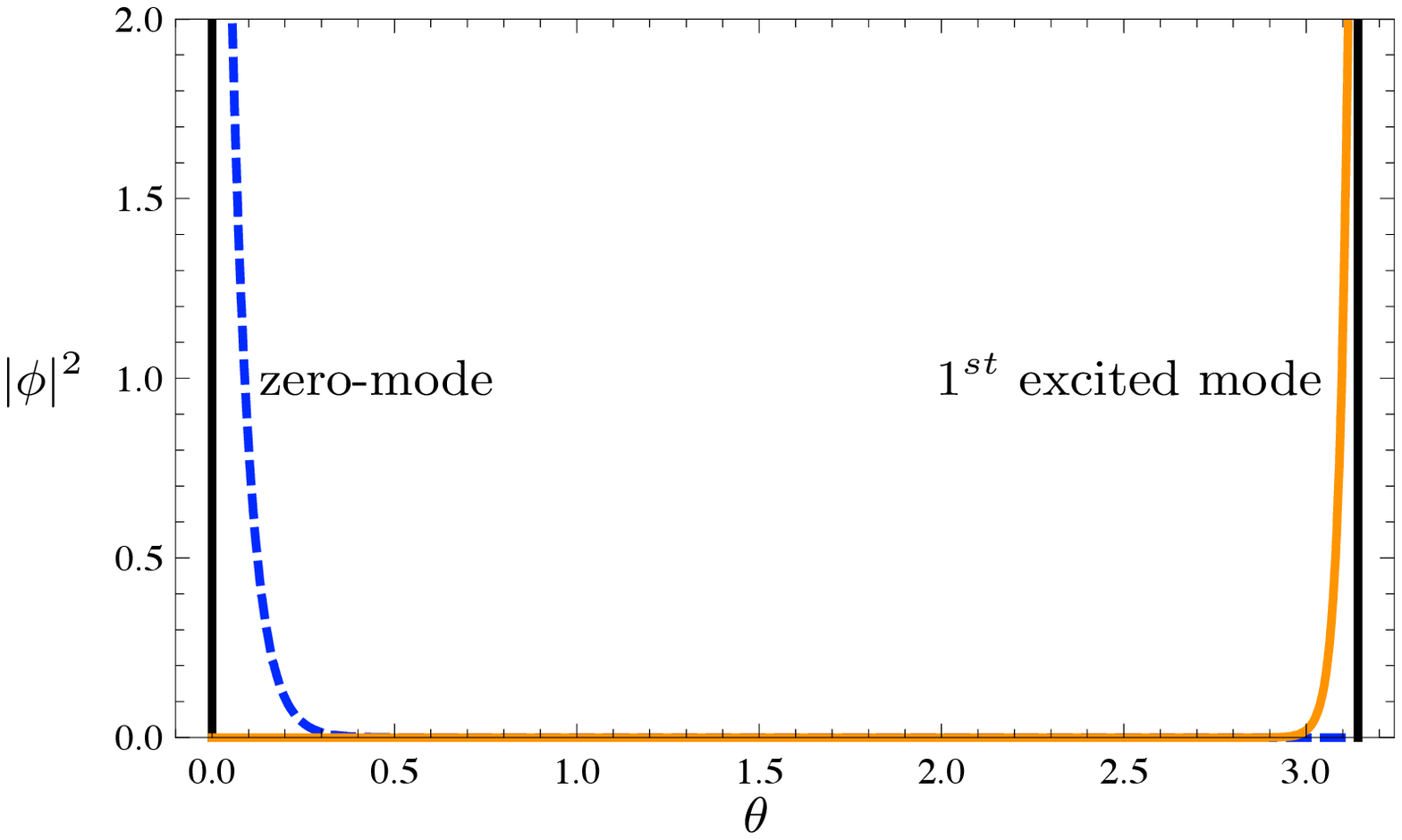}
   \caption{The figure illustrates, for the case of a massless minimally coupled bulk scalar field, the different localization properties of the massless, zero-mode (blue, dashed line), and the first excited mode (orange, continuous line). The branes are located at $\theta=0$ and $\theta=\pi$ and are represented by vertical, black continuous lines.}
\label{figura2}
\end{figure}

An interesting situation arises when the lowest mode of the Kaluza-Klein
tower is (nearly) massless but the model is still compatible with the 
observations because its wave function has a small overlap with 
the matter on the visible brane. 
Here, to illustrate such situation, 
we can think of the Randall-Sundrum I model (RS) \cite{rs},
where two, respectively positive and negative tension $3-$branes, bound
a slice of five-dimensional anti-de Sitter space. The mass scales of
the model are comparable to some Planckian cut-off scale, $M_P$, but
energy scales measured on the visible, negative tension brane, are redshifted by
a geometrical warp factor, $a\equiv e^{-\pi kr_c}$. Here $k\leq M_P$ is
the inverse of the AdS curvature scale and $r_c$ is the width of the
slice. By taking $kr_c \approx 12$, masses on the negative tension 
brane are of order $aM_P\sim$TeV. 
Let us consider a massless and minimally coupled scalar field propagating 
in the bulk with untwisted boundary conditions. 
Then, the wave function can be decomposed as before leading to
the standard tower-like effective action, but the four-dimensional 
spectrum contains a normalizable, massless zero-mode. 
Such mode is localized on the hidden brane side 
and its wave function overlap with matter localized on the visible brane is
small. The localization properties of a massless, minimally coupled
scalar field propagating in the RS two brane model are
illustrated in Fig.~\ref{figura2} for the massless mode and the first
massive excitation.
In such a case we should treat the boundary conditions more carefully 
since extending the plates into the bulk completely neglects  
the important effect of suppression owing to the small wave-function overlap. 
Namely, the boundary 
conditions have to be imposed on the field value on the brane. 
Contrary to four dimensional fields, which are
constrained to remain within the plates (in the idealized case), 
a bulk field can propagate to the other side of the plate via
the bulk. The situation is illustrated in Fig.~\ref{figura}-b. 
This means that, even in the case of ideal plates, partial penetration is
still possible via the bulk. Furthermore, the boundary condition should
be imposed not on each mode but on the field $\Phi$ in total:
\begin{equation}
 \Phi|_{\mbox{\scriptsize on the plates}}=0.\label{bcndts}
\end{equation}
In section IV, we will address this case. 

\section{Casimir effect with higher dimensional plates.} 

In this section we will analyze the contribution to the Casimir effect
from a higher dimensional field when the plates are artificially extended into the extra dimensions. 
We also assume that the 
coupling of $\varphi$ to ordinary matter localized on the brane is
so strong that the plates perfectly reflect the field, 
We do not take into account the back-reaction of this field on the
geometry. 
The masses of the modes in the Kaluza-Klein tower are given by $m_n$
with $n=1, 2, 3\cdots$.  
In the present discussion the details of the mass spectrum are
irrelevant. 

The Casimir energy per unit area is given by 
$$
{\cal E}=
\sum_{n=1}^\infty
 {\cal E}_{n}
 \equiv \sum_{n=1}^\infty \left[\lim_{L\to \infty}
 \lim_{s\rightarrow 0}{\hbar c\over 2 L^2} \zeta_n(s)\right]~,
$$
with 
\begin{widetext}
\beq
\zeta_n(s) \equiv \mu^{s} \sum_{n_1=1}^\infty 
\sum_{n_2=-\infty}^\infty \sum_{n_3=-\infty}^\infty 
\left[\left({\pi n_1\over \ell}\right)^2+
\left({2\pi n_2\over L}\right)^2+\left({2\pi n_3\over L}\right)^2
+m_n^2\right]^{(1-s)/2}~.
\eeq
The constant $\mu$ is a renormalization scale. 
Taking the limit $L \rightarrow \infty$, 
summations over $n_2$ and $n_3$ are replaced with integrals. 
Performing these integrals, we get
\beq
{\cal E}_{n}=
\lim_{s\rightarrow 0}-{\hbar c \mu^{s}\pi^{2-s} \over 4(3-s)\ell^{3-s}} 
\sum_{n_1=1}^\infty 
\left[ n_1^2+{m_n^2 \ell^2\over \pi^2}
\right]^{-(s-3)/2}~. 
\label{cas}
\eeq
The summation over $n_1$ can be obtained 
using a simplified version of 
the Chowla-Selberg formula 
(see for example \cite{cs,kk,ee}), 
\begin{eqnarray}
{\cal E}_{n}=
\lim_{s\rightarrow 0}{\hbar c \mu^{s}m_n^{3-s} \over 8\pi(3-s)} 
 \left[1-{\Gamma(-2+{s\over 2}) m_n \ell \over \sqrt{\pi} \Gamma(-{3\over 2}+{s\over 2})}
- {4(m_n \ell)^{-1+{s\over 2}}\over \pi \Gamma({3\over 2}+{s\over 2})}
 \sum_{p=1}^\infty p^{-2+{s\over 2}}K_{-2+{s\over 2}}\left(2 p m_n
						      \ell\right)\right]~, 
\end{eqnarray}
\end{widetext}
where $K_n$ represents the standard $n$-th order modified Bessel function. 
Notice that 
the Casimir force is obtained by taking the derivative of the 
Casimir energy and reversing the sign. 
The first term in the square brackets is independent of the separation
 of the two plates $\ell$, and hence 
it does not contribute to the Casimir force. 
The second term is 
exactly linear in $\ell$. 
Adding the contribution from the field outside the two plates cancels exactly
the $\ell$ dependence of this term. Therefore the second 
term is also irrelevant for the Casimir force. 
Then, the only remaining piece is the last term. 

The argument of the modified Bessel function  
depends on two scales: the
 plate separation, $\ell$, and the Kaluza-Klein mass $m_n$. 
In typical Casimir effect experiments $\ell$ is a fraction of a micrometer, 
while the Kaluza-Klein mass is at least $\sim$TeV.  
Hence, we have $m_n \ell\agt 10^{13}$. In the large argument limit 
the modified Bessel function is expanded as $K_{n}(z)\approx \sqrt{\pi/2z}\,e^{-z}$. 
Then, the Casimir force per unit area from each mode is evaluated to
\begin{eqnarray}
{\cal P}_{n}=-{d{\cal E}_n\over d\ell}
\approx {\hbar c m_n^{5/2} \over 3\pi^2\ell^{3/2}}e^{-2m_n\ell}. 
\end{eqnarray}
In the above expression  
the regularization parameter $s$ has been already relaxed to 0. 
It is possible to take this limit before taking the summation over $n$ 
because it is manifest that 
this summation is not divergent unless the desity of states of
the mass spectrum increases exponentially fast. 
This result shows that the Casimir force is simply given by 
the sum of contributions from respective Kaluza-Klein modes
under the approximation in which the plates are 
assumed to be extended into the bulk. 

These approximate boundary conditions are expected to over-estimate the Casimir force. 
Nevertheless, the expected effect due to bulk fields is exponentially
suppressed. Thus, we can conclude that massive Kaluza-Klein modes do not 
give significant correction to the Casimir force by any chance.

\section{Localized Boundary Conditions}

In this section, we consider the case of a delocalized massless
 mode with an associated massive Kaluza-Klein tower. 
The RS two-brane model, with the SM fields localized on the visible brane \cite{rs}, is an example of this sort. 

\begin{figure}
\includegraphics[height=2cm]{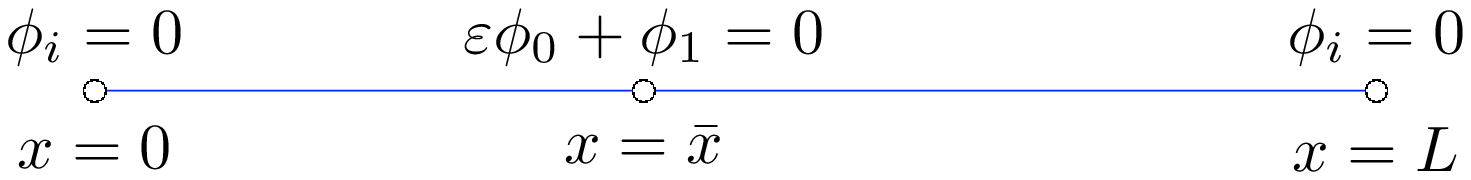}
   \caption{The figure illustrates the boundary conditions.}
\label{figura3}
\end{figure}

In general the problem is complicated due to the mixing between the various excitations. Here, in order to simplify the computation, we will consider only the contribution to the Casimir force from the massless zero mode and the first Kaluza-Klein excitation (As we have seen in the previous section, even 
over-estimating the Casimir force, the contribution of massive modes is still negligible). In principle, our method can be extended to include a finite number of excitations. For convenience we rewrite the four dimensional effective action composed of the above two terms:
\beq
S=-{1\over 2} \int d^4x \left\{
\phi_0 \Box \phi_0
+\phi_1 \left( \Box +m_1^2 \right) \phi_1
\right\}~.
\eeq
The plates are located at $x=0$, and $x=\ell$ 
and we will also introduce an artificial boundary at $x=L$. Such artificial boundary is sent to infinity at the end. The situation is described in Fig.~\ref{figura3}. 
The boundary condition (\ref{bcndts}) becomes
\bea
0&=&\phi_1+ \varepsilon \phi_0|_{x=\ell}~,\nonumber
\eea
while the perpendicular combination $\phi_0 - \varepsilon \phi_1$ must be smooth at $x=\ell$. 
The quantity $\varepsilon$ is a {\it small} constant that fixes the amplitude of the wave function at the brane.
Imposing Dirichlet boundary conditions at $x=0,~L$,
\bea
0&=&\phi_i|_{x=0,L}~,~~\mbox{for $i=0,1$}~,\nonumber
\eea
fixes the wave functions in the $x-$direction to be
\bea
\phi_0&=&a_1 \sin \left(\omega x\right)~,~~~~~~~~~~\mbox{for $0\leq x< \ell$}~,\nonumber\\ 
\phi_1&=&b_1 \sin \left(\kappa x\right)~,~~~~~~~~~~~\mbox{for $0\leq x< \ell$}~,\nonumber\\
\phi_0&=&a_2 \sin \left(\omega (x-L)\right)~,~~\mbox{for $\ell<x\leq L $}~,\nonumber\\ 
\phi_1&=&b_2 \sin \left(\kappa (x-L)\right)~,~~~\mbox{for $\ell< x\leq L$}~.\nonumber\\
\nonumber\eea
The boundary conditions can be arranged in matrix form leading to
\begin{widetext}
\bea
\left( \begin{array}{cccc}
\sin\left(\omega \ell\right) & -\varepsilon \sin\left(\kappa \ell\right)  & \sin\left(\omega \ell'\right) & -\varepsilon \sin\left(\kappa  \ell'\right) \\
\omega\cos\left(\omega \ell\right) & -\varepsilon \kappa \cos\left(\kappa \ell\right)  & -\omega\cos\left(\omega \ell'\right) & \varepsilon \kappa \cos\left(\kappa  \ell'\right) \\
\varepsilon \sin\left(\omega \ell\right) & \sin\left(\kappa \ell\right)  & 0 & 0 \\
0 & 0 & -\varepsilon \sin\left(\omega \ell'\right) & -\sin\left(\kappa \ell'\right)
\end{array} \right)
\left( \begin{array}{c}
a_1\\
b_1\\
a_2\\
b_2
\end{array} \right)=0~,
\eea
where $\ell'\equiv L-\ell$.
The quantization condition on the frequencies is then given by equating the determinant of the above matrix to zero:
\bea
0&=&\omega \sin\left(\kappa \ell\right) \sin\left(\kappa \ell'\right)\sin\left(\omega L\right)
+\varepsilon^2 \kappa \sin\left(\omega \ell\right) \sin\left(\omega \ell'\right)\sin\left(\kappa L\right)
~.
\nonumber
\eea
\end{widetext}
The above equation can be solved iteratively by expanding around the $\varepsilon=0$ solution as
\bea
\kappa&=&\kappa_0+\varepsilon^2 \kappa_1+...~,\label{left0}\\
\omega&=&\omega_0+\varepsilon^2 \omega_1+...~.\label{left1}
\eea
The lowest order eigenvalues are given by 
\begin{eqnarray*}
\kappa^{(1)}_0&=&{\pi n\over \ell},\\
\kappa^{(2)}_0&=&{\pi n\over \ell'},\\
\omega^{(3)}_0&=&{\pi n\over L},
\end{eqnarray*}
while next order corrections are given by 
\begin{eqnarray*}
\kappa^{(1)}_1&=&-{\kappa_0^{(1)}\over 2\omega^{(1)}_0 \ell}\sin(2\omega_0^{(1)}\ell),\\
\kappa^{(2)}_1&=&-{\kappa_0^{(2)}\over 2\omega^{(2)}_0 \ell'}\sin(2\omega_0^{(2)}\ell),\\
\omega^{(3)}_1&=&{\kappa_0^{(3)}\over 2\omega^{(3)}_0 L}
 \left(1-\cos(2\omega_0^{(3)}\ell)\right)\cot k_0 \ell.
\end{eqnarray*}
In the above expressions, we dropped terms that vanish after taking average over $L$ (keeping finite $L$ till the end will give the result for a piston localized on the brane). The above procedure can be automatically iterated and the corrections computed to any desired order. 
\subsection{Leading order corrections to the Casimir energy.}
The Casimir energy can be computed in the usual way. Expanding up to second order in $\varepsilon$ one arrives at the following expression:
\bea
{\cal E}&=&{\hbar c\mu^{s}\over 8\pi} {\Gamma((s-3)/2)\over \Gamma((s-1)/2)}\left( 
Z(s) + (3-s) \varepsilon^2 Y(s)\right)\nonumber\\
&+&O(\varepsilon^4)\label{zet}~,
\eea
with
\bea
Z(s)&\equiv& \sum_{}\omega_0^{(3-s)}~,\nonumber\\ 
Y(s)&\equiv&\sum_{} \omega_1 \omega_0^{(2-s)}~,\nonumber
\eea
where the sums extend over all the allowed configurations of eigenvalues given in the previous sub-section. 
The first term in (\ref{zet}) corresponds to the zero-order contribution. The only eigenvalue set that contributes to the Casimir energy in the large $L$ limit is $\left\{\kappa_0^{(1)},\kappa_1^{(1)}\right\}$. This is not surprising, since we expect the contribution from the massive mode to be the dominant one at leading order in $\varepsilon$. The explicit expression can be cast in the form of 
an Epstein-Hurwitz zeta function and rearranged by using the Chowla-Selberg formula as in the previous section.
Explicitly, the leading ($O(\varepsilon^0)$) correction to the energy reads
\bea
{\cal E}&=&-{\hbar c \over 8}{\lambda^2\over \ell^3} \sum_n K_{-2}\left(2\pi \lambda n\right)~,
\eea
where we have defined $\lambda \equiv m_1 \ell/\pi$. For large $\lambda$, the large argument expansion of the Bessel function leads again to an exponentially suppressed contribution to the energy,
\bea
{\cal E}&\simeq&-{\hbar c \over 16}{\lambda^{3/2}\over \ell^3} e^{-2\pi \lambda}~.
\eea
This is to be expected and it is in analogy with the discussion in conclusion to Section III. 
\begin{widetext}
\subsection{sub-leading corrections}
The sub-leading correction to the energy is instead more complicated to evaluate. We have to consider the contributions from the three sets of eigenvalues. The first contribution can be cast in the following form:
\begin{eqnarray}
Y^{(1)}&\equiv& \sum \omega_1^{(1)}(\omega_0^{(1)})^{2-s}\cr
 &=&-{1\over 2\ell}\left({\pi\over \ell}\right)^{2-s}
 \sum n^2 (n^2+\lambda^2)^{-s/2}\sin(2\pi\sqrt{n^2+\lambda^2})~.
\end{eqnarray}
The above sum, for some special values of $s$, can be recast in terms of Schl\"{o}milch-type series \cite{watson}. However, due to the fact that we need to analytically continue to $s=0$, we cannot directly apply the results available for those examples. The most direct way ia to use the Abel-Plana summation formula. However, as we will see, the above contribution will be exactely canceled by a contribution from $Y^{(3)}$ below, sparing us from its explicit computation. The contribution from the second set of eigenvalues is:
\begin{eqnarray}
Y^{(2)}
&\equiv& \sum \omega_1^{(2)}(\omega_0^{(2)})^{2-s}\cr
 &=&-{1\over 2\ell'}\left({\pi\over \ell'}\right)^{2-s}
 \sum n^2 (n^2+\lambda'{}^2)^{-s/2}\sin\left(2\pi{\ell\over \ell'}
       \sqrt{n^2+\lambda'{}^2}\right)~,
\end{eqnarray}
where $\lambda'\equiv m_1 \ell'/\pi$. In the large $L$ limit, the above sum can be recast in the form of an integral:
\begin{eqnarray}
Y^{(2)}
&\equiv& -{1\over 4\pi \ell} \left({1\over 2\ell}\right)^{2-s}
 \int_{\tilde\lambda}^\infty dt\,
 t^{1-s}(t^2-\tilde\lambda^2)^{1/2}\sin t\cr
 &=&{1\over 16\pi \ell^3} 
 \Re \left({i\over 2\ell}\right)^{-s}
 \int_{i\tilde\lambda}^\infty dy\,
 y^{1-s}(y^2+\tilde\lambda^2)^{1/2}
 e^{-y},  
\end{eqnarray}
where $\tilde\lambda\equiv 2\pi \lambda=2m_1\ell$. The above integral is finite for $s=0$, and
it can be easily evaluated expanding the integrand for large $\lambda$:
\begin{eqnarray}
Y^{(2)}
 = {1\over 32\sqrt{\pi} \ell^3}
 \left\{\cos\tilde\lambda
  \left(\tilde\lambda^{3/2}-{15\over 8}\tilde\lambda^{1/2}-{105\over 128}\tilde\lambda^{-1/2}+\cdots \right)
 +\sin\tilde\lambda
  \left(\tilde\lambda^{3/2}-{15\over 8}\tilde\lambda^{1/2}+{105\over 128}\tilde\lambda^{-1/2}+\cdots \right)
 \right\}. 
\end{eqnarray}
The contribution from the third set of eigenvalues leads to
\begin{eqnarray}
Y^{(3)}&\equiv& \sum \omega_1^{(3)}(\omega_0^{(3)})^{2-s}\cr
 &=&{1\over 2L}
 \sum \omega_0^{1-s}\sqrt{\omega_0^2+m^2}(1-\cos 2\omega_0\ell)\cot
 \left(\sqrt{\omega_0^2+m^2}\ell\right).
\end{eqnarray}
Setting $t=\sqrt{\omega_0^2+m^2}\ell/\pi$, and taking large $L$
 limit, the summation over $n$ is, again, replaced with the integral 
$\int dt\, \ell^2\omega_0/(\pi t L)$: 
\begin{eqnarray}
Y^{(3)}
 ={\pi^2\over 2 \ell^3} \left({\pi \over \ell}\right)^{-s}
 P\int_{0}^\infty dt\,
 t^{2}(t^2+\lambda^2)^{-s/2}\left(
   1-\cos(2\pi\sqrt{t^2+\lambda^2})\right)\cot \pi t~, 
\label{Y3}
\end{eqnarray}
where $P\int$ means the principal part integral.
We divide this expression into two parts. The first part can be written as
\begin{eqnarray}
Y^{(3)}_1
 & \equiv &{\pi^2\over 2 \ell^3} \left({\pi \over \ell}\right)^{-s}
 P\int_{0}^\infty dt\,
 t^{2}(t^2+\lambda^2)^{-s/2}\cot [\pi (t+i\epsilon)]\cr
 & = & -{\pi^2\over 2 \ell^3} \left({\pi \over \ell}\right)^{-s}
 \Re \left[\int_{0}^\infty du\,
 u^{2}(\lambda^2-u^2)^{-s/2}(\coth(\pi u)-1)
 +\int_{0}^\infty du\,
 u^{2}(\lambda^2-u^2)^{-s/2}\right]. 
\end{eqnarray}
The first integral in the last line of the above expression is finite for $s\to0$. Hence, setting $s=0$, we obtain 
\begin{eqnarray}
\Re \int_{0}^\infty du\,
 u^{2}(\coth(\pi u)-1)={\zeta(3)\over 2\pi^3}. 
\end{eqnarray}
The second integral can be evaluated by analytical continuation as
\begin{eqnarray}
\Re \int_{0}^\infty du\,
 u^{2}(\lambda^2-u^2)^{-s/2}
=-{\sqrt{\pi}\lambda^{3-s} e^{\pi i s/2}\Gamma[(s-3)/2]\over
4\Gamma[s/2]}
\to 0, \qquad \mbox{for }s\to 0. 
\end{eqnarray}
The second part of (\ref{Y3}) can be decomposed into the 
contribution from the contour integral and residues as 
\begin{eqnarray}
Y^{(3)}_2
 & \equiv & -{\pi^2\over 2 \ell^3} \left({\pi \over \ell}\right)^{-s}
 \Biggl[
  \Re \int_{0}^\infty dt\,
 t^{2}(t^2+\lambda^2)^{-s/2}\cot [\pi (t+i\epsilon)]
 \exp[2\pi i\sqrt{t^2+\lambda^2}]\cr
&&\hspace{2cm}
 -\sum 2\pi n^2(n^2+\lambda^2)^{-s/2} \sin(2\pi\sqrt{n^2+\lambda^2})
\Biggr]. \label{eqy32}
\end{eqnarray}
The part corresponding to the second term in the square brackets, 
which comes from the residues, completely cancels $Y^{(1)}$, as we anticipated. 
Hence, the combination $Y^{(1)}+Y^{(3)}_2$ is 
expressed by the part corresponding to the remaining term in 
the square brackets in (\ref{eqy32}). This contribution is finite 
in the limit $s\to 0$, and, rotating the phase of the integration contour by $\pi/2$, we have 
\begin{eqnarray}
Y^{(1)}+Y^{(3)}_2
 & = & {\pi^2\over 2 \ell^3} 
  \Re \int_{0}^\infty du\,
 u^{2} [(\coth(\pi u)-1)+1]
 \exp(2\pi i\sqrt{\lambda^2-u^2}). 
\end{eqnarray}
As for the second term in the square brackets, 
by changing the integration variable to $y=2\pi\sqrt{u^2-\lambda}$, 
we find that this contribution is exactly the same as $Y^{(2)}$. 
The first term can be evaluated by expanding with respect to $\tilde \lambda$, 
leading to 
\begin{eqnarray}
&&{\pi^2\over 2 \ell^3} 
\int_{0}^\infty du\,
 u^{2}  \exp(2\pi i\sqrt{\lambda^2-u^2})(\coth(\pi u)-1)\cr
&&\qquad ={1\over 8 \pi \ell^3} \Biggl\{
 \cos(\tilde \lambda)\left(\zeta(3)+{45\zeta(7)\over \tilde\lambda^2}
  -{1260\zeta(9)-4725\zeta(11)\over\tilde\lambda^4}+\cdots\right)\cr
&&\qquad\qquad\qquad
 +\sin(\tilde \lambda)\left({6\zeta(5)\over \tilde\lambda}
  +{45\zeta(7)-420\zeta(9)\over\tilde\lambda^3}+\cdots\right)
\Biggr\}. 
\end{eqnarray}
Combining all results, we can write the $O(\varepsilon^2)$ correction to the Casimir energy, $\delta{\cal E}$, as
\begin{eqnarray}
\delta{\cal E}&=& \varepsilon^2 {\hbar c \over 16\pi^2 \ell^3}
 \Biggl\{
\zeta(3)-
\cos(\tilde\lambda)
\left[
\sqrt{\pi}
\left(\tilde\lambda^{3/2}-{15\over 8}\tilde\lambda^{1/2}-{105\over 128}\tilde\lambda^{-1/2}+\cdots \right)
+\left(-\zeta(3)+{45\zeta(7)\over \tilde\lambda^2}+\cdots\right)
\right]
\cr
&&\qquad 
-\sin(\tilde\lambda)
\left[
\sqrt{\pi}
\left(\tilde\lambda^{3/2}-{15\over 8}\tilde\lambda^{1/2}+{105\over 128}\tilde\lambda^{-1/2}+\cdots \right)
+\left({6\zeta(5)\over \tilde\lambda}+{45\zeta(7)-420\zeta(9)\over\tilde\lambda^3}+\cdots\right)\right]
\Biggr\}.
\end{eqnarray}
\end{widetext}
From the above expression it is clear that there is no exponential suppression for the corrections of $O(\varepsilon^2)$, and the only suppression comes from the fact that $\varepsilon$ is small if the wave function overlap of the zero mode with matter on the visible brane is small. 

\section{Constraints on new physics from Casimir force experiments}
Before concluding our paper, we wish to summarize our point of view concerning the role of the corrections to the Casimir force that can in principle be produced by additional degrees of freedom to those of the SM. 
In considering modifications to the Casimir force and the possibility that these may be detected with forthcoming precision experiments, it is necessary to clarify that the model itself is not ruled out {\it a priori}. 
When extra degrees of freedom (regardless whether they come from higher dimensions or arise in different ways) are introduced, in order for Casimir force experiments to be able to reaveal their presence, these extra fields must be coupled to the SM and the SM must be charged under these additional degrees of freedom. Clearly, the possibility that the SM particles are strongly coupled to the above additional fields is not admissible. 
A first essential point is that the Casimir force arises as a quantum effect, and thus it is sub-leading with respect to the classical long range force arising from massless modes. In order for such quantum contributions to become dominant, the long range force must be suppressed. One possibility is that SM particles are very weakly coupled to the extra degree(s) of freedom, not excluding the possibility that the Casimir force, which arises as a collective effect, is non negligible. Notice that standard boundary conditions used in Casimir force calculations can be a reasonable approximation even if the coupling is weak. The above arguments lead various people to compute the Casimir force under the presence of extra spatial dimensions. The general conclusion was that the corrections from higher dimensional degrees of freedom are sizeable.

In this work we reconsidered the above issues and focused our attention to the non-trivial case of RS-type models, in which the weak coupling between the higher dimensional degrees of freedom and the SM localized on the brane arises as an effect of small wave function overlap. Contrary to the above mentioned claims, our results show that the corrections because of higher dimensional degrees of freedom are negligible. Our method does not rely on the details of the model. It applies whenever weak coupling is realized due to small wave function overlap, offering one possible (correct) way to compute the force. 

In conclusion to this section, we wish to mention another important point: whether the results we have obtained for the RS model, are generic. One possible alternative way to realize the suppression of the classical long range force is to consider two plates uncharged under the additional degrees of freedom. In this case, although the plates are uncharged, atoms should be assumed to possess an additional dipole moment. The presence of dipoles realizes Dirichlet-type boundary conditions for the additional degrees of freedom and generate a Casimir-like force. However, in order to avoid conflict with observations, one must forbid  un-paired monopole charges. This implies that, at least at low energies, these extra degrees of freedom must be confining. 
How Casimir force experiments may constrain such confining theories is a very interesting question, but beyond the scope of our paper. We hope to return to this issue in the near future. 

\section{Conclusions}
In this paper we analyzed the corrections to the Casimir energy between two perfectly conducting parallel plates located on a brane embedded in a higher dimensional space. We clarified the role of the boundary conditions at the plates and pointed out that imposing the boundary conditions individually on each Kaluza-Klein mode corresponds to artificially extending the plates in the bulk. This produces in general an over-estimation of the Casimir force. 
In this case, we explicitly showed that the correction to the Casimir effect is exponentially suppressed, if all the Kaluza-Klein modes are massive. The contribution from the zero mode is more subtle to examine. First of all, the presence of a massless mode localized on the visible brane, strongly coupled to matter on the brane, must be excluded for obvious reasons. A less obvious case, is when the massless mode is not localized on the visible brane. This case cannot be ruled out from the beginning if the coupling between the plates and the extra massless mode is small, even if SM fields (and thus the plates) are charged under this extra degree of freedom. In this case, production of this extra mode would be suppressed, for example, in particle collision, however, we cannot rule out the possibility that observable effects may arise as a collective phenomena like the Casimir effect. We described a method to study the Casimir effect, when the zero-mode has a small wave function overlap compared to the massive excitations. We introduced a new form of boundary conditions that is imposed not on each Kaluza-Klein excitation individually, but on a linear combination of zero mode and massive excitations. When the amplitude of the zero mode wave function on the visible brane is small, the Casimir energy can be computed perturbatively to any desired order. We carried out this computation to second order and showed that bulk fields only induce small corrections to the Casimir effect. The suppression arises at zero order in the relative localization between the massless and massive modes, described by a parameter $\varepsilon$, because of the fact that the zero mode is completely decoupled and the contribution comes only from the massive excitation, which is exponentially suppressed. At next order in $\varepsilon$, the correction is only suppressed (by the factor $\varepsilon^2$) due to small wave function overlap. 
The above results seem to indicate that it is difficult to get stringent constraints on models of RS type, in its simplest incarnations, from Casimir force measurements.
Generalization of our results can be easily obtained for different types of bulk field and are expected to hold when the relative amplitude of the zero mode wave function on the visible brane with respect to the massive mode(s) is small.

\acknowledgements
We wish to thank M. Minamitsuji and S. Mukohyama for various discussions on the subject.
This work is supported by the JSPS through Grants Nos. 19GS0219, 20740133, 19540285, 21244033. 
The support of the Global COE Program ``The Next Generation of Physics, Spun from Universality and Emergence''
from the Ministry of Education, Culture, Sports, Science and Technology (MEXT) of Japan is kindly acknowledged. 
We also acknowledge the Organization of the GCOE/YITP workshop (YITP-W-09-01) on `Non-linear cosmological perturbations'.

\end{document}